\documentclass[twoside,twocolumn,9pt]{article}
\usepackage{extsizes}
\usepackage[super,sort&compress,comma]{natbib} 
\usepackage[version=3]{mhchem}
\usepackage[left=1.5cm, right=1.5cm, top=1.785cm, bottom=2.0cm]{geometry}
\usepackage{balance}
\usepackage{mathptmx}
\usepackage{sectsty}
\usepackage{graphicx} 
\usepackage{lastpage}
\usepackage[format=plain,justification=justified,singlelinecheck=false,font={stretch=1.125,small,sf},labelfont=bf,labelsep=space]{caption}
\usepackage{float}
\usepackage{fancyhdr}
\usepackage{fnpos}
\usepackage[english]{babel}
\addto{\captionsenglish}{%
  
}
\usepackage{array}
\usepackage{droidsans}
\usepackage{charter}
\usepackage[T1]{fontenc}
\usepackage[usenames,dvipsnames]{xcolor}
\usepackage{setspace}
\usepackage[compact]{titlesec}
\usepackage{hyperref}
\usepackage{bm}
\usepackage[switch]{lineno}

\usepackage{epstopdf}

\definecolor{cream}{RGB}{222,217,201}

\begin{document}
\pagestyle{fancy}
\thispagestyle{plain}
\fancypagestyle{plain}{
\renewcommand{\headrulewidth}{0pt}
}

\makeFNbottom
\makeatletter
\renewcommand\LARGE{\@setfontsize\LARGE{15pt}{17}}
\renewcommand\Large{\@setfontsize\Large{12pt}{14}}
\renewcommand\large{\@setfontsize\large{10pt}{12}}
\renewcommand\footnotesize{\@setfontsize\footnotesize{7pt}{10}}
\makeatother

\renewcommand{\thefootnote}{\fnsymbol{footnote}}
\renewcommand\footnoterule{\vspace*{1pt}%
\color{cream}\hrule width 3.5in height 0.4pt \color{black}\vspace*{5pt}} 
\setcounter{secnumdepth}{5}

\makeatletter 
\renewcommand\@biblabel[1]{#1}            
\renewcommand\@makefntext[1]%
{\noindent\makebox[0pt][r]{\@thefnmark\,}#1}
\makeatother 
\renewcommand{\figurename}{\small{Fig.}~}
\sectionfont{\sffamily\Large}
\subsectionfont{\normalsize}
\subsubsectionfont{\bf}
\setstretch{1.125} 
\setlength{\skip\footins}{0.8cm}
\setlength{\footnotesep}{0.25cm}
\setlength{\jot}{10pt}
\titlespacing*{\section}{0pt}{4pt}{4pt}
\titlespacing*{\subsection}{0pt}{15pt}{1pt}

\fancyfoot{}
\fancyfoot[LO,RE]{\vspace{-7.1pt}\includegraphics[height=9pt]{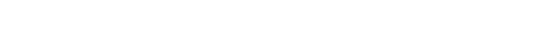}}
\fancyfoot[CO]{\vspace{-7.1pt}\hspace{13.2cm}\includegraphics{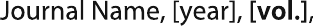}}
\fancyfoot[CE]{\vspace{-7.2pt}\hspace{-14.2cm}\includegraphics{head_foot/RF}}
\fancyfoot[RO]{\footnotesize{\sffamily{1--\pageref{LastPage} ~\textbar  \hspace{2pt}\thepage}}}
\fancyfoot[LE]{\footnotesize{\sffamily{\thepage~\textbar\hspace{3.45cm} 1--\pageref{LastPage}}}}
\fancyhead{}
\renewcommand{\headrulewidth}{0pt} 
\renewcommand{\footrulewidth}{0pt}
\setlength{\arrayrulewidth}{1pt}
\setlength{\columnsep}{6.5mm}
\setlength\bibsep{1pt}

\makeatletter 
\newlength{\figrulesep} 
\setlength{\figrulesep}{0.5\textfloatsep} 

\newcommand{\topfigrule}{\vspace*{-1pt}%
\noindent{\color{cream}\rule[-\figrulesep]{\columnwidth}{1.5pt}} }

\newcommand{\botfigrule}{\vspace*{-2pt}%
\noindent{\color{cream}\rule[\figrulesep]{\columnwidth}{1.5pt}} }

\newcommand{\dblfigrule}{\vspace*{-1pt}%
\noindent{\color{cream}\rule[-\figrulesep]{\textwidth}{1.5pt}} }

\makeatother

\twocolumn[
  \begin{@twocolumnfalse}
{\includegraphics[height=30pt]{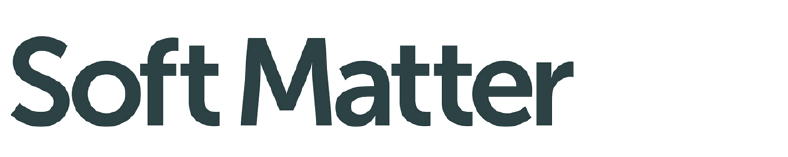}\hfill\raisebox{0pt}[0pt][0pt]{\includegraphics[height=55pt]{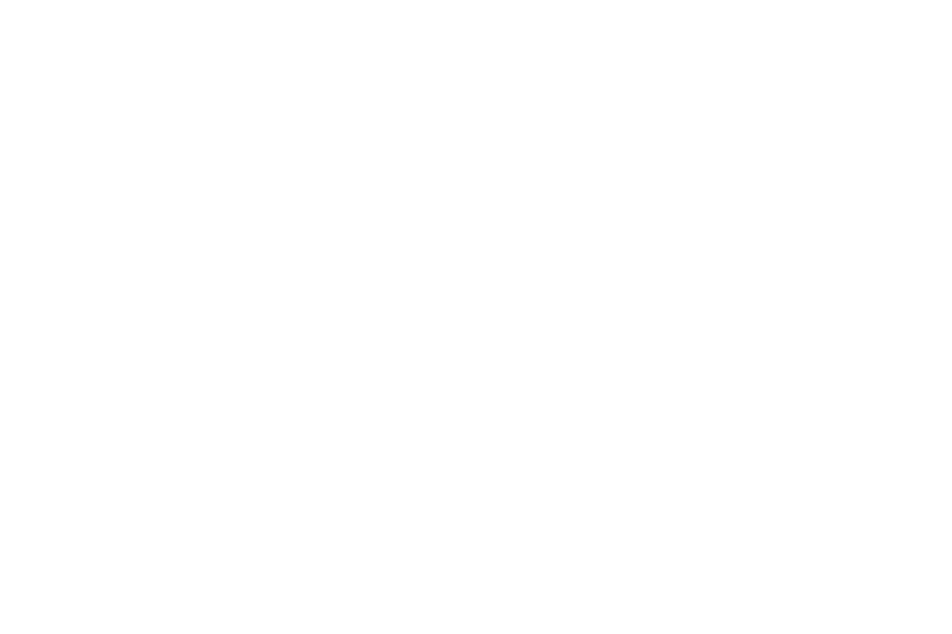}}\\[1ex]
\includegraphics[width=18.5cm]{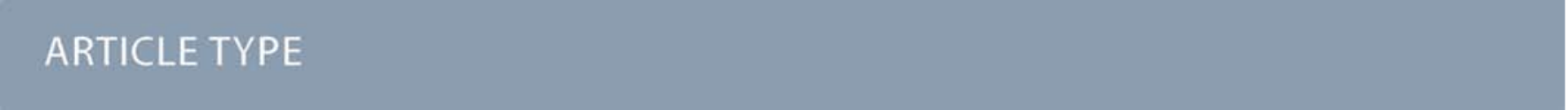}}\par
\vspace{1em}
\sffamily
\begin{tabular}{m{4.5cm} p{13.5cm} }

\includegraphics{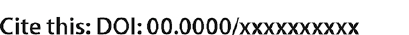} & \noindent\LARGE{\textbf{Theory of polymer diffusion in polymer-nanoparticle mixtures: effect of nanoparticle concentration and polymer length}} \\
\vspace{0.3cm} & \vspace{0.3cm} \\

 & \noindent\large{Bokai Zhang,$^{\ast}$\textit{$^{a}$},Jian Li\textit{$^{b}$}, Juanmei Hu,\textit{$^{a}$} and Lei Liu$^{\ast}$\textit{$^{a}$}} \\

\includegraphics{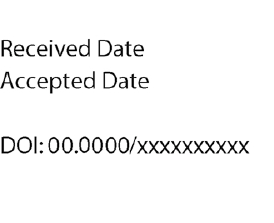} & \noindent\normalsize{The dynamics of polymer-nanoparticle (NP) mixtures, which involves multiple scales and system-specific variables, has posed a long-standing challenge on its theoretical description.  In this paper, we construct a microscopic theory for polymer diffusion in the mixtures based on a combination of generalized Langevin equation, mode-coupling approach, and polymer physics ideas.  The parameter-free theory has an explicit expression and remains tractable on pair correlation level with system-specific equilibrium structures as input.  Taking a minimal polymer-NP mixture as an example, our theory correctly captures the dependence of polymer diffusion on NP concentration and average interparticle distance.  Importantly, the polymer diffusion exhibits a power law decay as the polymer length increases at dense NPs and/or long chain, which marks the emergence of entanglement-like motion.  The work provides a first-principle theoretical foundation to investigate dynamic problems in diverse polymer nanocomposites.} \\

\end{tabular}

 \end{@twocolumnfalse} \vspace{0.6cm}

  ]

\renewcommand*\rmdefault{bch}\normalfont\upshape
\rmfamily
\section*{}
\vspace{-1cm}


\footnotetext{\textit{$^{a}$Department of Physics, Zhejiang Sci-Tech University, Hangzhou 310018, China}}
\footnotetext{\textit{$^{b}$Department of Physics and Electronic Engineering, Heze University, Heze 274015, China}}
\footnotetext{\textit{$\ast$ Corresponding Author: Bokai Zhang, E-mail: bkzhang@zstu.edu.cn;  Lei Liu, E-mail: leiliu@zstu.edu.cn}}



\section{Introduction}
Understanding the diffusion of flexible polymer in polymer-nanoparticle (NP) mixtures is an important problem as it is widely related to transport of biopolymer, \cite{PNAS2007,NatMater2009,AnnRevBio2008} drug delivery \cite{Brigger2012,NBio2003} and the properties and processability of polymer nanocomposites (PNCs) materials. \cite{Kumar2017,Kumar2017JCP}  The description of dynamics in polymer-NP mixtures involves multiple system-specific factors, including the characteristics of NP (e.g., sizes, shapes, interface softness, and concentration) and polymer (e.g., topology, molecular weight, and rigidity) and various polymer-NP interactions.  Thus, a theory for polymer diffusion on these factors represents a tough challenge and poses a complex problem, where the motion of polymer and NP is interrelated, thus significantly influencing each other.  

In the polymer-NP mixtures, NP diameter, volume fraction and polymer chain length are basic variables to control equilibrium structure, relaxation and transport.  Their effect on various properties of the mixtures has attracted an enormous amount of research using computer simulations and experimental approaches as well as some phenomenological models. \cite{BAILEY2020,Colby2014,Holt2013,Richter2013PRL2,Choi2013,Gam2011,Echeverria2010,Karatrantos2017,Meth2013,Richter2011,Sorichetti2018,Xie2017}  Elastic recoil detection (ERD) experiments measuring the diffusion of polymer tracers in athermal NPs reveal remarkable reduction compared to their diffusion in the bulk. \cite{Choi2013,Gam2011}  A universal relation between polymer diffusion and the average interparticle distance (ID) normalized by the radius of gyration, $h/R_g$, has been reported, suggesting that $h/R_g$ captures the effect of NP size, size polydispersity, and concentration on polymer dynamics. \cite{Choi2013} Further, in recent simulations, a scaling law relating polymer diffusion to $h/R_g$ was also found.\cite{Sorichetti2018,Luo2019,Patti2017}  The functional dependence of the diffusion on the NP size polydispersity is proposed in a simple theoretical framework. \cite{Patti2017}

Polymer length is a key quantity in polymer materials and plays an important role in chain dynamics.  For instance, there is an entanglement crossover in linear polymer melts as polymerization increases beyond a certain threshold $N_e$. \cite{McLeish2002} The polymer chain length $N_p$-scaling of polymer diffusion changes from $D_p\sim N_p^{-1}$ to $D_p\sim N_p^{-2}$ .  For ring polymer melts and solutions, long chains facilitate topological threading and clustering, consequently hinder chain motion due to their connectivity and uncrossability. \cite{Michieletto2017,EPLZhang2020} Besides, chain stiffness also has a significant impact on transport and dynamics.  For instance, the local chain dynamics influenced by the stiffness was found to couple with the motion of NPs.\cite{Yethiraj2011}  The stiffness-induced long-range nematic order determines the dynamics in PNCs. \cite{Yethiraj1998,Patti20172}

In the polymer-NP mixtures, the polymer length has a significant impact on dynamics, in particular, for the high size ratio of polymer to NP, i.e., the radius of gyration greater than NP diameter, $R_g/\sigma_n>1$.  In recent years, Small-angle X-ray scattering (SAXS) experiment, molecular dynamics simulations, and nonlinear Langevin equation theory found, in comparison with large NPs, small NPs ($R_g>\sigma_n$) can induce a more pronounced reduction in polymer diffusion, greater variation in glass transition temperature, and higher dynamic fragility. \cite{Sorichetti2018,Xie2017,Li2014}   Large scale molecular dynamics simulations of entangled polymer melts reported that small NPs  slow down the polymer diffusion by 40\% at most. In addition, the simulations also revealed that the NP-segment size ratio has a significant influence on polymer diffusion. \cite{Kalathi2014} Furthermore, as recently found in experiment, chain motion in PNCs shows a crossover from entanglement caused by polymers to that caused by the NPs.  Chains diffuse in a tube formed by NPs rather than surrounding chains. \cite{Richter2011}  More recently, Cao \emph{et al}. found that the concentration of NP in unentangled polymers induces chain reptation-like motion reflected by the stress relaxation modulus. \cite{Cao2019}

In theory, Asakura-Oosawa model, which regards polymer as a soft sphere with effective pair potential, largely captures equilibrium and dynamic properties of real polymer-NP mixtures in the so-called "colloidal" limit, where polymers are much smaller than NPs, $2R_g \ll\sigma_n$. \cite{AOmodel}  But in "protein" limit ($2R_g\geq \sigma_n$), polymer diffusion theory is still quite limited.  Meth \emph{et al}. developed a phenomenological theory where polymer diffusion in the network formed by PNCs is mapped to the motion of spheres through hollow cylinders. \cite{Meth2013}  Following this picture, an analytical expression for the polymer diffusion is obtained and provides several predictions that qualitatively and quantitatively consistent with experiments.  But the approach is oversimplified as important detailed information, e.g., chain conformation and microscopic interaction, is neglected. 

In this paper, our goal is to develop a microscopic, parameter-free and testable theory of long-time center-of-mass (CM) diffusion of polymer influenced by NP additions, including the effect of NP volume fraction and polymer chain length.  An explicit expression for the diffusion coefficient is derived based on the expression of resistance in generalized Langevin equation and naive mode-coupling (naive-MCT) approximation \cite{Kirkpatrick1987} which has been successfully applied to NPs diffusion in PNCs, \cite{Yamamoto2011} activated dynamics in binary mixtures \cite{Jadrich2012,Zhang2018} and cross-linked networks. \cite{Dell2014}  To demonstrate the applicability of our theory, we calculated the polymer diffusion in a minimal polymer-NP mixture model that ignores possible interaction between polymers. \cite{Fuchs2001}  We use the minimal mixture model for three main reasons.  (i) Both experiment and simulation studies reveal that various properties of PNCs are mainly influenced and controlled by interfacial phase behavior and chain motion around the NPs. \cite{Roland2016, Sokolov2020, Li2014}  Thus, comprehending the relationship between local structure and mobility at the level of force between the single polymer and NPs is the first step to understanding chain dynamics in more complicated PNCs materials. 
(ii) The essential features in polymer-NP interaction can be captured by this model.  It provides a proper description of the interaction between the nonadsorbing polymer and hard sphere and the chain conformational change around NPs.  
(iii) The simplified model has analytical and explicit expressions of structural correlations, which are needed as input functions, and renders the theory tractable.

The paper is organized as follows: section \ref{sec:2} presents a general integral equation theory available to solve equilibrium structures, the description of a minimal polymer-NP mixture model, and the main steps of the theoretical derivation of the long-time polymer CM diffusion.  In section \ref{sec:3}, we present the numerical results for polymer diffusion.  The effect of NP volume fraction and polymer chain length is mainly focused on.  Qualitative and quantitative comparisons with simulation and experiment are discussed.  Finally, summary and discussion are presented in  section \ref{sec:4}, including the application possibilities, limitations and the future development of the theory.  Explicit expressions and main steps for derivation are contained in Appendics.

\section{Model and theory}\label{sec:2}
\subsection{Equilibrium structure}
The polymer reference interaction site model (PRISM) approach is an integral equation theory applied to solve intermolecular pair equilibrium structure correlations for liquid-like polymer
systems. \cite{Schweizer1997}  In the PRISM approach, the segments of polymers and NPs are modeled as a series of interaction sites.  The relation between the site-site intermolecular total correlation function $\hat{\bm{H}}(\bm{k})$ and intermolecular direct correlation function (DCF) $\hat{\bm{C}}(\bm{k})$ is described by the well-known Ornstein-Zernicke (OZ) equation
\begin{equation}\label{Eq1}
	\begin{aligned}
		  \hat{\bm{H}}(\bm{k})=\hat{\bm{\Omega}}(\bm{k})\hat{\bm{C}}(\bm{k})[\hat{\bm{\Omega}}(\bm{k})+\hat{\bm{H}}(\bm{k})]
	\end{aligned}
\end{equation}
where $\hat{\bm{\Omega}}(\bm{k})$ is the intramolecular structure factor matrix in which the diagonal terms $\hat{\Omega}_{ii}=\rho_i \hat{\omega}_i$ and the cross terms $\hat{\Omega}_{ij}=(\rho_i+\rho_j) \hat{\omega}_{ij}$.  Here we set $N_i$ as the site number of type $i$ and $\rho_i=N_i/V$ as the site number density of type $i$.  In the NP-polymer mixture, the site type index $i$ of polymer segment and NP is labeled by the subscript $p$ and $n$, respectively.  For chemical homogeneous sites, the only nonzero terms in the intramolecular structure factor matrix are $\hat{\omega}_{p}$ for chain and $\hat{\omega}_{n}=1$ for NP.  $\hat{H}_{ij}(k)$ and $\hat{C}_{ij}(k)$ are the Fourier transforms of $\rho_i \rho_j h_{ij} (r)$ and $C_{ij}(r)$, respectively. 

In real space, the total correlation function $h_{ij}$ between site $i$ and $j$ is related to radial distribution function $g_{ij}$ by
\begin{equation}\label{Eq2}
	\begin{aligned}
  h_{ij}(r)=g_{ij}(r)-1
	\end{aligned}
\end{equation}
Here, the hardcore condition prevents these sites from overlapping, and then radial distribution function follows
\begin{equation}\label{Eq3}
  g_{ij} (r<\frac{1}{2} (\sigma_i+\sigma_j ))=0
\end{equation}

To solve the OZ equation, an extra closure for DCF is necessary.  In liquid theory, DCF $C_{ij}$ represents interaction between site $i$ and $j$.  In PNCs, real force between polymer segments and NPs can be renormalized as an effective force, $\bm{F}_{eff}=-k_BT \bm{\nabla} C_{np}(\bm{r})$.  Under random phase approximation (RPA), DCF is taken as an overlapping core and relevant to real potential by $C_{ij}^{RPA}=-V_{ij} (\bm{r})/k_B T$.  For higher order approximation, DCF for nonoverlapping Percus-Yevick (PY) closure in isotropic media is taken as
\begin{equation}\label{Eq4}
C_{ij}^{PY} (r)=\left\{
\begin{aligned}
 &0 &r>\frac{1}{2} \big(\sigma_i+\sigma_j \big)  \\
 &-e^{V_{ij}(r)/k_BT}g(r)& r\le\frac{1}{2} \big(\sigma_i+\sigma_j \big)
\end{aligned}
\right.
\end{equation}
With the site number density and intramolecular structure factor as input, the intermolecular pair correlation can be solved numerically by combining eqn  (\ref{Eq1})-(\ref{Eq4}).
\begin{figure}[!t]
	\centering{\includegraphics[width=.45\textwidth]{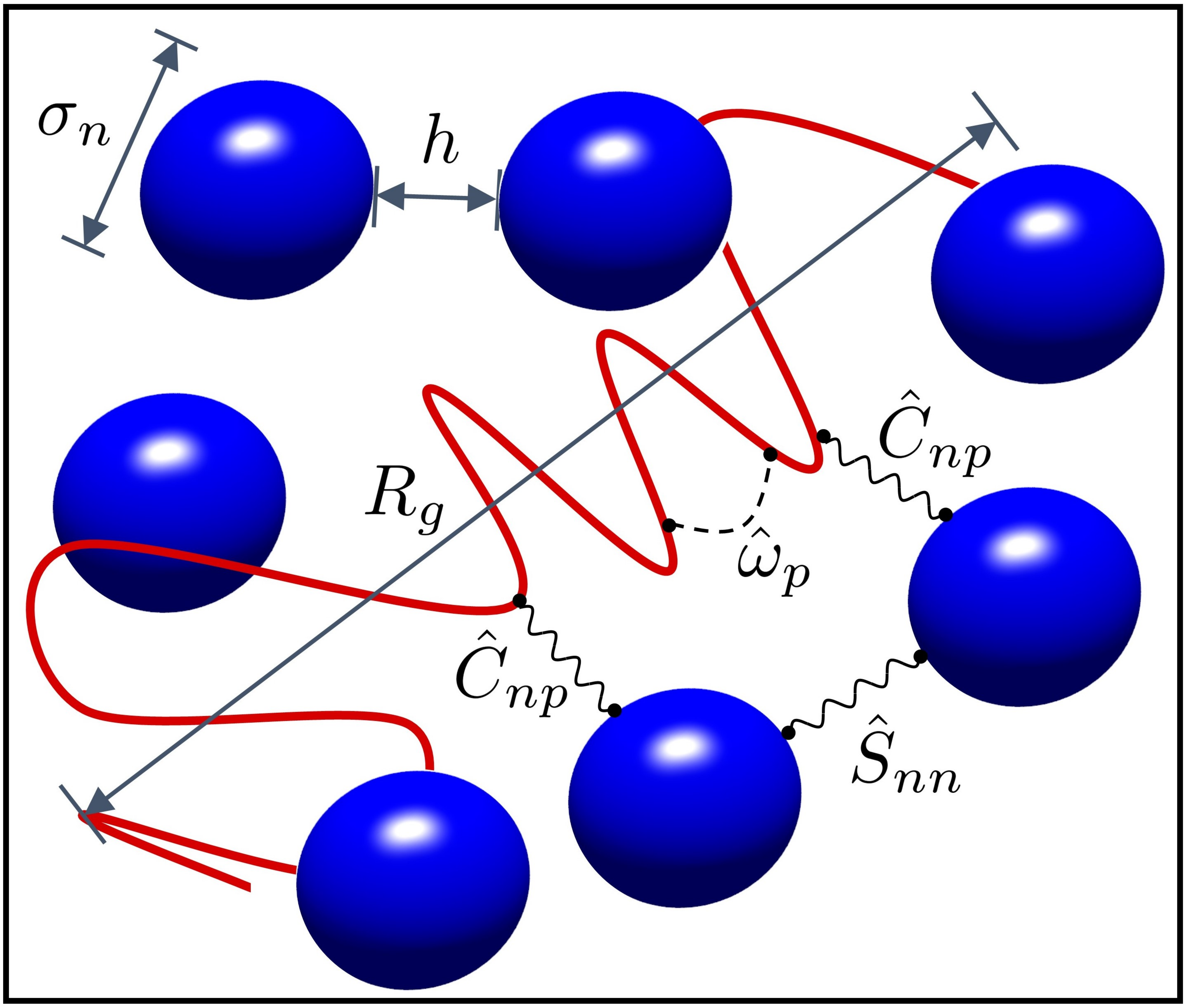}}
	\caption{Schematic diagram of the minimal polymer-NP mixture model. A single polymer chain (red curve) diffusing in NPs (blue spheres) indicating key length scales and relevant structure correlations in the force-force vertex function, $\hat{V}$, eqn (\ref{Eq14}).  Wavy lines represent intermolecular interactions, direct correlation function $\hat{C}_{np}$ and static strcuture factor $\hat{S}_{nn}$, respectively. Dashed line represents intramolecular structure factor $\hat{\omega}_p$.}
	\label{fig:1}
\end{figure}

\subsection{Polymer-NP mixture model}
We consider a binary mixture of hard spheres with diameter $\sigma_n$ and nonadsorbing flexible polymers with the radius of gyration $R_g$ as illustrated in Fig. \ref{fig:1}.  In a minimal polymer-NP mixture model, the limit of dilute polymer is taken, namely, the number density of polymer chain is taken as $\rho_p\to 0$. \cite{Fuchs2001}  Polymer is modeled as a Gaussian thread chain with hard-sphere beads as statistical segments.  As illustrated by the red curve in Fig. \ref{fig:1}, a point-like approximation is adopted (the bead diameter $\sigma_p\to 0$), which is expected to be accurate when the NP is much larger than the segment diameter.  The intramolecular structure factor is taken as
\begin{equation}\label{Eq5}
	\begin{aligned}
		\hat{\omega}_{p} (\bm{k})=\frac{N_p}{1+\bm{k}^2 R_g^2/2}
	\end{aligned}
\end{equation} 
Considering the change of polymer conformation close to the surface of the NPs, a modified version of PY (mPY) closure including the entropic repulsive range $\lambda$ is proposed, \cite{Fuchs2001}
\begin{equation}\label{Eq6}
  C_{np}^{mPY} (\bm{r})=\frac{1}{4\pi\lambda^2} \int  \frac{1}{|\bm{r}-\bm{s}|}   e^{|\bm{r}-\bm{s}|/\lambda} C_{np}^{PY} (\bm{s}) d\bm{s}
\end{equation}
In $k$ space, eqn (\ref{Eq6}) can be transformed to a Lorentzian-like form,
\begin{equation}\label{Eq7}
  \hat{C}_{np}^{mPY}  (\bm{k})=\frac{\hat{C}_{np}^{PY} (\bm{k})}{1+\bm{k}^2 \lambda^2}
\end{equation}
The DCF in PY closure, $C_{np}^{PY}$, incorporates the interaction between unconnected polymer segments and NPs.  The spatial convolution in eqn (\ref{Eq6}) captures the chain conformation change and medium-range nonlocal interaction as segment is close to the surface of the NP.  Consequently, in the modified DCF, the entropic repulsive range $\lambda$ is considered to incorporate chain connectivity and conformational entropic effect.  The undetermined distance $\lambda$ modifies rapid variations of DCF around NPs and makes them smoother.  The magnitude of $\lambda$ is estimated as the order of or smaller than the polymer persistence length and NP diameter.  

Following a thermodynamic consistency method by computing insertion free energy and free volume, the interaction distance $\lambda$ can be obtained. \cite{Fuchs2001}  In the limit of dilute polymers, $\lambda$ as functions of NPs volume fraction, $\phi_n$, and polymer radius of gyration, $R_g$, can be approximately expressed as 
\begin{equation}\label{Eq8}
  \lambda^{-1}=\frac{\sqrt{2}}{R_g}+\frac{1+2\phi_n}{1-\phi_n}\frac{4}{\sigma_n(\sqrt{5}-1)}
\end{equation}
In general, static structure of the mixtures with repulsive or attractive interactions can be solved numerically by employing simple Picard iteration or inexact Newton method.\cite{Lisa2008} For the mixtures with pure hard-sphere interactions, an analytical and explicit expression for the DCF is possible. Substituting the above formula into eqn (\ref{Eq1}) and combining with the modified PY closure eqn (\ref{Eq7}), isotropic DCF in $k$ space can be derived as
\begin{equation}\label{Eq9}
\begin{aligned}
\hat{C}_{np}(k)=&\frac{\sigma_p^2}{\sigma_n^2}\Big\{\hat{Q}_{np}(k)\Big(1+i k \xi_0\Big)\Big(1+i k \lambda\Big)\\
+&e^{-ik/2}\Big[1-\hat{Q}_{nn}(k)\Big]2\pi\Big(u_b+ik\lambda v_b\Big)\Big\}
\end{aligned}
\end{equation}
where $\xi_0=R_g/\sqrt{2}$ is the polymer persistence length.  These functions, $\hat{Q}_{np}$, $\hat{Q}_{nn}$, $u_b$, and $v_b$, depend on $R_g$ and $\phi_n$. Their complete expressions can be found in Appendix A.  

\subsection{Theory of polymer center-of-mass diffusion}
We first formulate the long-time resistance of tagged chain on the CM level based on Mori-Zwanzig projection operator technique and naive-MCT factorization approximation, \cite{Zwanzig, Gotze2008} which can effectively incorporate friction arising from surrounding environment.  Starting point of the derivation is the generalized Langevin equation for the CM position of polymer, $\bm{R}_i$, \cite{Hansen2013} 
\begin{equation}\label{Eq10}
  \xi_s \frac{d\bm{R}_i}{dt}=-\int_{-\infty}^t d\tau K_{CM} (t-\tau)\frac{d\bm{R}_i(\tau)}{d\tau}+\bm{f}^Q (t)
\end{equation}
where $\xi_s$ is short time friction constant and $\bm{f}^Q$ is a random fluctuating force obeying the fluctuation-dissipation theorem.  The memory frictional kernel $K_{CM}$ captures the viscoelastic effect.  It is related to the force-force time correlation function of the tagged polymer by 
\begin{equation}\label{Eq11}
K_{CM}(t)=\frac{1}{3k_BT}\langle \bm{F}_{CM}(0)\cdot\bm{F}_{CM}(t)\rangle
\end{equation}
where $\bm{F}_{CM}=\sum_i^{N_p} \bm{F}_i$ is the total force exerted on a chain .  The resistance is obtained by integrating the memory kernel over time, $\xi_{CM}=\int_0^\infty  K_{CM}(t)dt$.  

According to the standard closure of naive-MCT, the total force is projected on a bilinear product of tagged chain density and collective density fluctuation, $\hat{b}_s(\bm{k},\bm{k}')= \delta\hat{p}_{T,p}(\bm{k}) \delta\hat{c}_s(\bm{k})$.  And then four-point correlation is approximately factorized into pair correlations.  More details can be found in Appendix B.  Finally, the force-force correlation function in isotropic media for multiple components is obtained
\begin{equation}\label{Eq12}
\begin{aligned}
  \langle \bm{F}_{CM}(0)\cdot\bm{F}_{CM}(t)\rangle&=\frac{N_{p}}{2\pi^2\beta^2}\int_0^{\infty}dkk^4\hat{\omega}_p(k) \\
&\times \sum_{j,m}\hat{C}_{pj}(k)\hat{S}_{jm}(k,t)\hat{C}_{mp}(k)\hat{\Gamma}_p^s(k,t)
\end{aligned}
\end{equation}
Here $\beta=1/k_BT$ and dynamical frozen model is adopted, $\hat{\omega}_p(\bm{k},t)=\hat{\omega}_p(\bm{k})\hat{\Gamma}_p^s(\bm{k},t)$. $\hat{S}_{jm}$ is collective dynamic structure factor and $\hat{\Gamma}_p^s$ is normalized dynamic correlation propagator for tagged chain.  In the polymer-NP binary mixture, the summation index $j$ and $m$ is over $n$ (NP) and $p$ (polymer).  

The resistance for the CM of polymer using the above summation in eqn (\ref{Eq12}) contains two separate contributions arising from polymer-NP and polymer-polymer interactions, $\xi_{CM}=\xi_{pn}+\xi_{pp}$.  $\xi_{pp}$ describes  the frictional effect of both direct interchain interaction and indirect coupling interaction mediated by NPs.  These interactions are expected to be negligible in dilute polymer solution or a single chain diffusing in the NP solutions where polymer chains have little overlap.  Considering the limit of dilute polymer $\rho_p\to 0$, the polymer-polymer frictional term $\xi_{pp}$ vanishes.  The final expression for the long-time resistance yields
\begin{equation}\label{Eq13}
	\begin{aligned}
		\xi_{CM}&=\frac{\rho_n}{6\pi^2\beta}\int_0^{\infty}dk\int_0^{\infty}dt\hat{V}(k)\hat{\Gamma}_{nn}^c(k,t)\hat{\Gamma}_p^s(k,t)
	\end{aligned}
\end{equation}
As illustrated in Fig. \ref{fig:1}, the static part of force-force vertex function $\hat{V}(k)$ represents time independent contribution to the resistance involving interactions at different length scales, including chain length $N_p$, DCF between NP and segment $\hat{C}_{np}$, static structure factor for NPs $\hat{S}_{nn}$, and intramolecular structure factor $\hat{\omega}_p$,
\begin{equation}\label{Eq14}
	\begin{aligned}
		\hat{V}(k)=N_pk^4\hat{\omega}_p(k)\hat{C}^2_{np}(k)\hat{S}_{nn}(k)
	\end{aligned}
\end{equation}

The time dependent part is the force relaxation channel via density correlation propagator of the tagged polymer, 
\begin{equation}\label{Eqb1}
\hat{\Gamma}_p^s(k,t)=\frac{1}{N_p}\sum_{i,j}^{N_p}\langle exp({i\bm{k}\cdot(\bm{R}_i^s(t)-\bm{R}_j^s(0))})\rangle
\end{equation}
and collective density correlation propagator of the NPs, 
\begin{equation}\label{Eqb2}
\hat{\Gamma}_{nn}^c(k,t)=\frac{1}{N_n}\sum_{i,j}^{N_n}\langle exp({i\bm{k}\cdot(\bm{R}_i^n(t)-\bm{R}_j^n(0))})\rangle
\end{equation}  
where $\bm{R}_i^s$ is the position of the i-th segment belonging to the tagged chain.  $\bm{R}_i^n$ is the position of the NP.  There are no exact expressions to $\hat{\Gamma}_p^s$ and $\hat{\Gamma}_{nn}^c$.  Some simple expressions for these propagators are adopted to make our theory analytically tractable.  At large length scale ($kR_g\ll 1$), A fast exponential decay on short time is appropriate, $\hat{\Gamma}_p^s\approx e^{-k^2 D_0t/N_p}$, where $D_{0}=k_BT/3\pi \sigma_p\eta_0$ is segment diffusion constant and $\eta_0$ is the solvent viscosity.  In the intermediate length scale, the internal motion of a single chain has an impact on the tagged-polymer density correlation propagator.  A dynamic random phase approximation (RPA) capturing length scale-dependent internal conformation and polymeric fractals is proposed,\cite{Muthukumar1985,Schweizer2020}
\begin{equation}\label{Eq15}
\hat{\Gamma}_p^s(k,t)=e^{-k^2D_{0}t/\hat{\omega}_p(k)}
\end{equation}

In MCT, the relaxation rate of density fluctuation is considered to correlate with length scale and be proportional to $1/\hat{S}(k)$ due to de Gennes narrowing effect.  For simplicity, a vineyard-like approximation at short times is applied. \cite{Schweizer2003} The collective density propagator for the NPs is taken as
\begin{equation}\label{Eq16}
\Gamma_{nn}^c(k,t)=e^{-{k^2D_{n0}t}/\hat{S}_{nn}(k)}
\end{equation}
where $D_{n0}={k_BT}/3\pi\eta_0\sigma_n$ is NP diffusion constant.  After integration over time, the resistance becomes
\begin{equation}\label{Eq17}
	\xi_{CM}=\frac{N_p\rho_n k_B T}{6\pi^2}\int_0^{\infty}\frac{k^2\hat{\omega}_p(k)\hat{C}_{np}^2(k)\hat{S}_{nn}(k)}{\frac{k_BT}{6\pi\eta_0 \sigma_p\hat{\omega}_p(k)}+\frac{k_BT}{6\pi\eta_0\sigma_n\hat{S}_{nn}}}dk
\end{equation}

Next, the long-time diffusion coefficient is related to the resistance by the well-known Einstein relation, $D_{p}=k_BT/\xi_{p}$.  We assume the total resistance $\xi_p$ contains two parts, $\xi_{p}=\xi_{bulk}+\xi_{CM}$.  $\xi_{bulk}$ is the resistance for single chain diffusing in the bulk without adding NPs.  It can be obtained based on the Rouse model , $\xi_{bulk}=3\pi \sigma_p N_p\eta_0$. \cite{Doi1988}  Considering $\sigma_n/\sigma_p\to \infty$ for Gaussian thread model, final expression for the polymer diffusion coefficient normalized by its value in the bulk, $D_{p0}=k_BT/\xi_{bulk}$, is explicitly derived as
\begin{equation}\label{Eq18}
	\begin{aligned}
		D_r&=\frac{D_{p}}{D_{p0}}&=&\frac{1}{1+\frac{\rho_n}{6\pi^2}\int_0^{\infty} dk k^2 \hat{\omega}_p^2\hat{C}^2_{np}\hat{S}_{nn}}
	\end{aligned}
\end{equation}
\begin{figure*}[!h]
	\includegraphics[width=.98\textwidth]{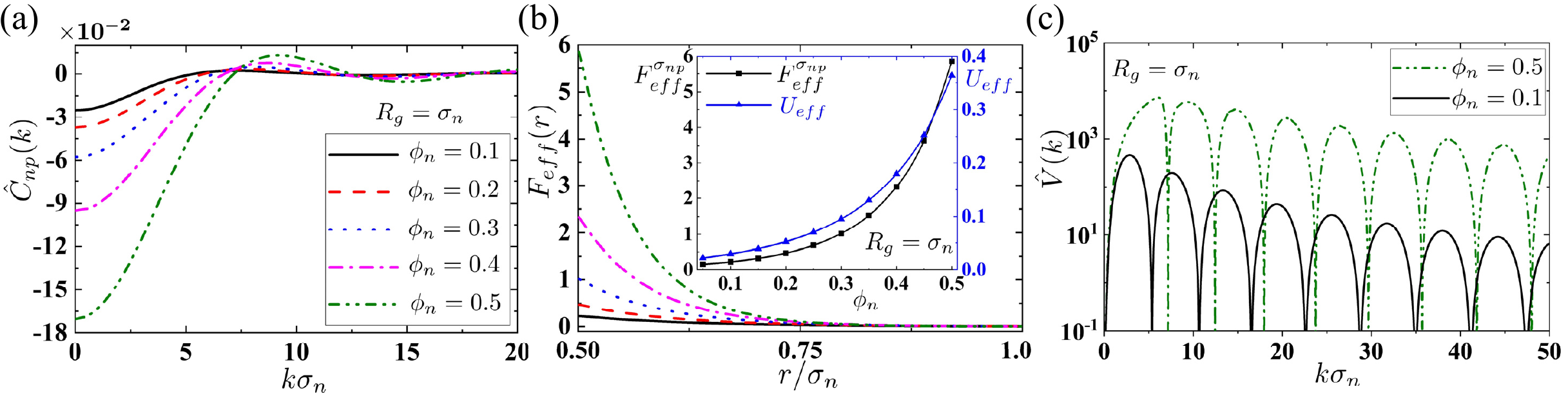}
	\caption{(a) Plots of DCF versus wavevector for several NP volume fractions at $R_g=\sigma_n$.  (b) segment-NP distance dependence of effective force. The legend is the same as in Figure (a).  Inset: The contact value of effective force (black square line) and EFW (blue triangle line) versus NP volume fraction.  (c) Static part of force-force vertex function for two NP volume fractions, $\phi_n=0.1$ and $\phi_n=0.5$.}
	\label{fig:2}
\end{figure*}
We adopt $\sigma_n/\sigma_p =10$ in the calculation of equilibrium structure throughout this study, given that the NP diameter $\sigma_n$ is much greater than segment diameter $\sigma_p$.  In the minimal polymer-NP mixture model, structure and dynamics of the polymer-NP mixture are controlled by two key variables: NP volume fraction, $\phi_n=\pi\rho_n\sigma_n^3/6$, and polymer-NP size ratio, $R_g/\sigma_n$.  The units of length and energy are respectively $[L]=\sigma_n$ and $[E]=k_BT$.  

\section{Results}\label{sec:3}
\subsection{NP concentration effect}
We first focus on NP concentration effect on equilibrium structure and effective interaction.  
Fig. \ref{fig:2}a presents representative calculations of wavevector-dependent DCF for various NP volume fractions at fixed chain length $R_g=\sigma_n$ using eqn (\ref{Eq9}).  Our results show that the concentration of NPs strongly influences behavior of the DCF, especially at small wavevector.  Quantitatively, when $\phi_n$ increases from $0.1$ to $0.5$, the magnitude of the DCF at $k\to 0$ increases nearly seven times, from $|\hat{C}_{np}(k\to 0)|\approx 0.025$ to $|\hat{C}_{np}(k\to 0)|\approx 0.170$.  The behavior of $\hat{C}(k)$ around zero wavevector is considered to determine large-scale physical properties, such as thermodynamics and phase separation boundary in polymer melts and PNCs. \cite{Chen2007,Lisa2008}  Moreover, there is also a significant increase in the height of primary peak in $\hat{C}_{np}$ with increasing of $\phi_n$, corresponding to more compact local packing between NPs and segments.  The peak position, $k_{max}$, is closely related to the mean nearest-neighbor distance between segment and NP, which is estimated as $r_{mean}\sim 2\pi/k_{max}$.  Fig. \ref{fig:2}a shows that the peak position $k_{max}$ has a shift with adding NPs, $k_{max}\approx 7.6-9.2\sigma_n^{-1}$, that is, $r_{mean}\approx 0.83-0.68\sigma_n$.  

In real space, real force exerted by surrounding particles can be averaged over fluctuations and renormalized as the effective force between NP and segment, which is written as $\bm{F}_{eff}^{np}(r)=-k_BT \bm{\nabla} C_{np}(r)$.  The quantity is considered to characterize local segment-NP interaction.  Fig. \ref{fig:2}b shows that the effective force as a function of the segment-NP distance reduced by the diameter of NP, $r/\sigma_n$.  The value of the effective force decays monotonically, from contact value at $r=\sigma_{np}\equiv(\sigma_n+\sigma_p)/2$ to zero at $r/\sigma_n>1$, indicating local segment-NP interaction range less than NP diameter.  As illustrated in Fig. \ref{fig:2}b, the positive contact value of the effective force around $\sigma_{np}$ indicates the presence of repulsive interaction between hard NP and nonadsorbing segment.  The magnitude of the effective force exhibits a significant increase with NP volume fraction $\phi_n$.  Specifically, the contact value increases, $F_{eff}^{\sigma_{np}}\approx 0.16-5.85$ with $\phi_n=0.05-0.5$, as illustrated by the black line in the inset of Fig. \ref{fig:2}b.

To further quantify the frictional effect arising from NPs, a renormalized effective frictional work (EFW) is introduced as $U_{eff}=\int_{\sigma_{np}}^{\infty} \bm{F}_{eff} (\bm{r})\cdot d\bm{r}$.  Physically, $U_{eff}$ equals to the work of effective force driving the NP-segment distance from $\sigma_{np}$ to infinity.  Due to the definition of the effective force, the EFW exactly equals to negative contact value of DCF in real space excerpt an energy unit $k_BT$, $U_{eff}=-k_BT C_{np}(\sigma_{np})$.  
The inset of Fig. \ref{fig:2}b shows that the EFW strongly increases with NP volume fraction more than one order of magnitude, $U_{eff}\approx 0.023-0.363$ with $\phi_n=0.05-0.5$.  

Fig. \ref{fig:2}c shows the representative results of the vertex function given in eqn (\ref{Eq14}) as a function of the nondimensional wavevector, $k\sigma_n$, for two different NP volume fractions.  We find that the concentration of NPs leads to a significant increase in the amplitude of the vertex function by one to two orders of magnitude, and thus significantly affects the resistance coefficient through the integral in eqn (\ref{Eq13}).

\begin{figure}[!t]
	\includegraphics[width=.49\textwidth]{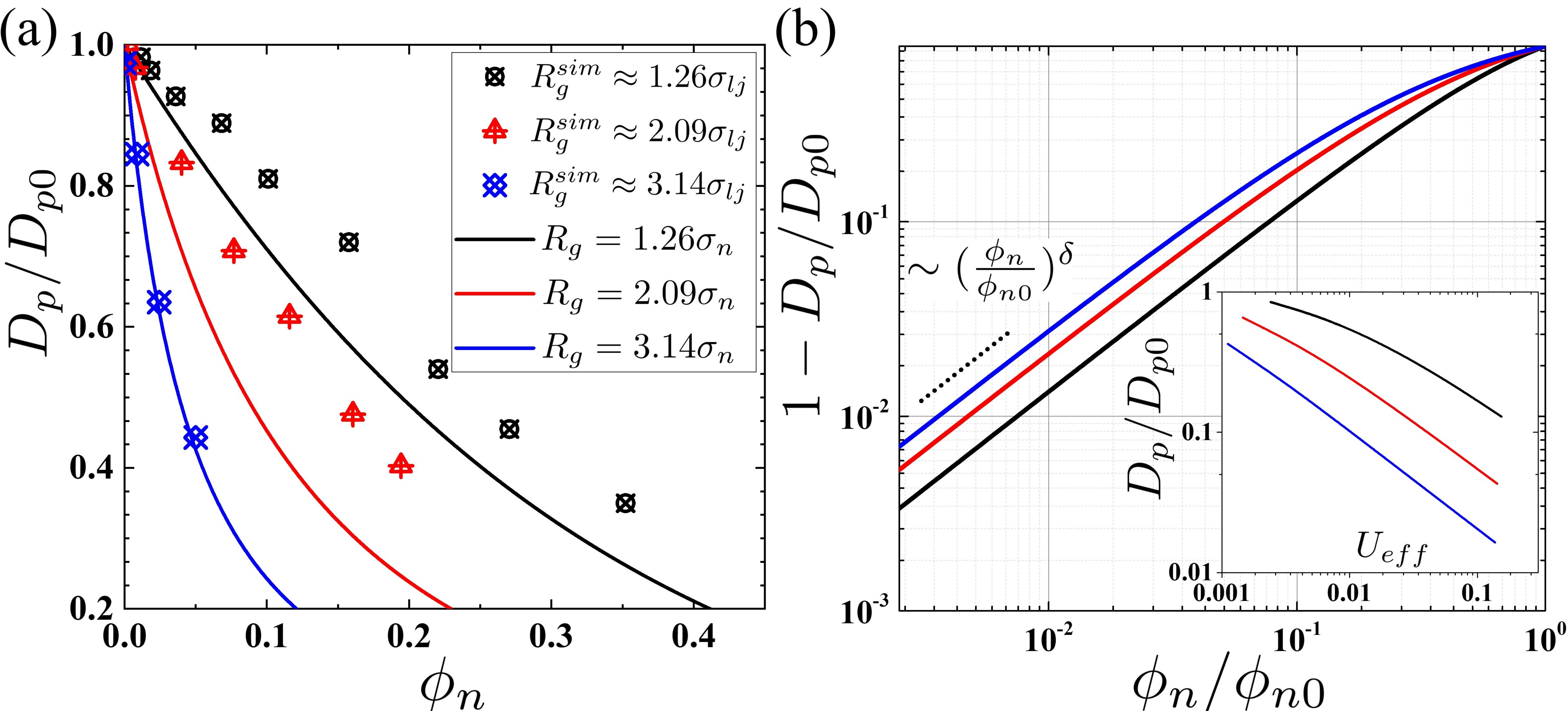}
	\caption{(a) Normalized polymer diffusion, $D_p/D_{p0}$ versus $\phi_n$ for several chain lengths computed from our theory (solid lines).  Simulation results are also plotted (symbols, ref. \citenum{Sorichetti2018}), where the radius of gyration changes less than $8\%$ and thus is assumed as a constant here.  (b) Plots of $1-D_p/D_{p0}$ versus $\phi_n$ rescaled by the cutoff volume fraction $\phi_{n0}$, where $D_p(\phi_{n0})/D_{p0}=0.2$.  The legend is the same as in Figure (a).  The dotted line represents power-law behavior described by eqn (\ref{Eq19}). Inset: Normalized polymer diffusion versus EFW. }
	\label{fig:3}
\end{figure}

We now qualitatively and quantitatively compare our theoretical predictions for polymer diffusion with recent results in simulations and experiments. \cite{Sorichetti2018, Gam2011} 
We restrict our comparison to those work in unentangled polymer systems, i.e., short chains or low concentration, due to the neglect for interactions between polymers in the minimal model.  Fig. \ref{fig:3}a presents the numerical calculations of the normalized polymer diffusion with adding NPs at different radii of gyration.  A rapid decay in the normalized diffusion is found as NP volume fraction increases.  The polymer diffusion exhibits a more drastic decrease for longer chain length.  For instance, $D_p$ decays to $20\%$ of $D_{p0}$ at $\phi_n\approx 0.12$ for $R_g=3.14\sigma_n$, whereas it decays to the same value at $\phi_n\approx 0.41$ for shorter chain $R_g=1.26\sigma_n$.

The symbols in Fig. \ref{fig:3}a show recent simulation results obtained from MD simulations of a mixture of Lennard-Jones NPs and FENE polymers as a function of $\phi_n$ for different chain lengths. \cite{Sorichetti2018}   In the simulation, the maximum monomer volume fraction is restricted to 0.148, which is much less than the critical entangled volume fraction, thus the system remains in the unentangled regime.  The polymer-polymer interactions are considered to be a minor factor for the diffusion.  Here we focus on the change of polymer diffusion influenced by the NPs.  The fast decay of the normalized diffusion with adding NPs and more dramatic change for longer chain is observed.  The polymer length dependence of the $D_r$-$\phi_n$ curve in simluation is qualitatively in agreement with our theoretical predictions.  For chain length $R_g=3.14\sigma_n$, the quantitative agreement is excellent for all NP volume fractions studied.  For shorter chain, $R_g = 2.09\sigma_n$ and $R_g=1.26\sigma_n$, the $D_r$-$\phi_n$ curves decrease more slowly compared to the theoretical results.  We emphasize that precise comparisons are subtle given the presence of three main differences in the simulation from our minimal model: (i) finite polymer density fraction effect due to greater monomer volume fraction in the simulation than the critical overlap volume fraction $\phi_m^*$.  (ii) weak attraction between polymers and NPs.  (iii) Some work has revealed that the ratio $\sigma_n/\sigma_p$ has a significant effect on polymer dynamics,\cite{Kalathi2014} which is neglected by the zero-thickness Gaussian thread model.

The decline degree of polymer diffusion due to NP concentration is quantified by $1-D_p/D_{p0}$ .  An empirical function related the quantity to $\phi_n$ is proposed \cite{Sorichetti2018}
\begin{equation}\label{Eq19}
	1-\frac{D_p}{D_{p0}}=\Big(\frac{\phi_n}{\phi_{n0}}\Big)^\delta
\end{equation}
The cutoff volume fraction $\phi_{n0}$ represents the location where arrest of motion of polymers occurs.  In practice, the value of the cutoff volume fraction can be set as $D_r(\phi_{n0})=0.2$ given that the minimum decay rate of the normalized diffusion measured in most experiments and simulations approaches to 20\% due to the limited accuracy. \cite{Li2014, Choi2013, Sorichetti2018}  The set value of cutoff volume fraction $\phi_{n0}$ does not greatly affect the behavior of the decayed diffusion.  Fig. \ref{fig:3}b shows our theoretical predictions for $1-D_p/D_{p0}$ as a function of NP volume fraction rescaled by $\phi_{n0}$.  We find that eqn (\ref{Eq19}) does a good description for the normalized diffusion at a small amount of NPs concentration.  The scaling exponent $\delta$ is approximately equal to 1 for all radii of gyration we studied.  As a comparison, $\delta$ obtained from simulations is about $0.76-1.15$ for various NP sizes.  For higher NP volume fraction, $\phi_n/\phi_{n0}>0.1$, the polymer diffusion deviates from the behavior described by eqn (\ref{Eq19}).  

We further compute the dependence of the diffusion on the EFW to illustrate how effective friction between NP and segment influences the polymer diffusion.  As shown in the inset of Fig. \ref{fig:3}b, polymer diffusion exhibits a significant decrease with EFW, suggesting that the concentration of NPs enhances effective frictional interaction mainly through contact interaction characterized by the DCF, $C_{np}(\sigma_{np})$, and thus suppresses the motion of polymer chain.

\begin{figure}[!t]
	\includegraphics[width=.48\textwidth]{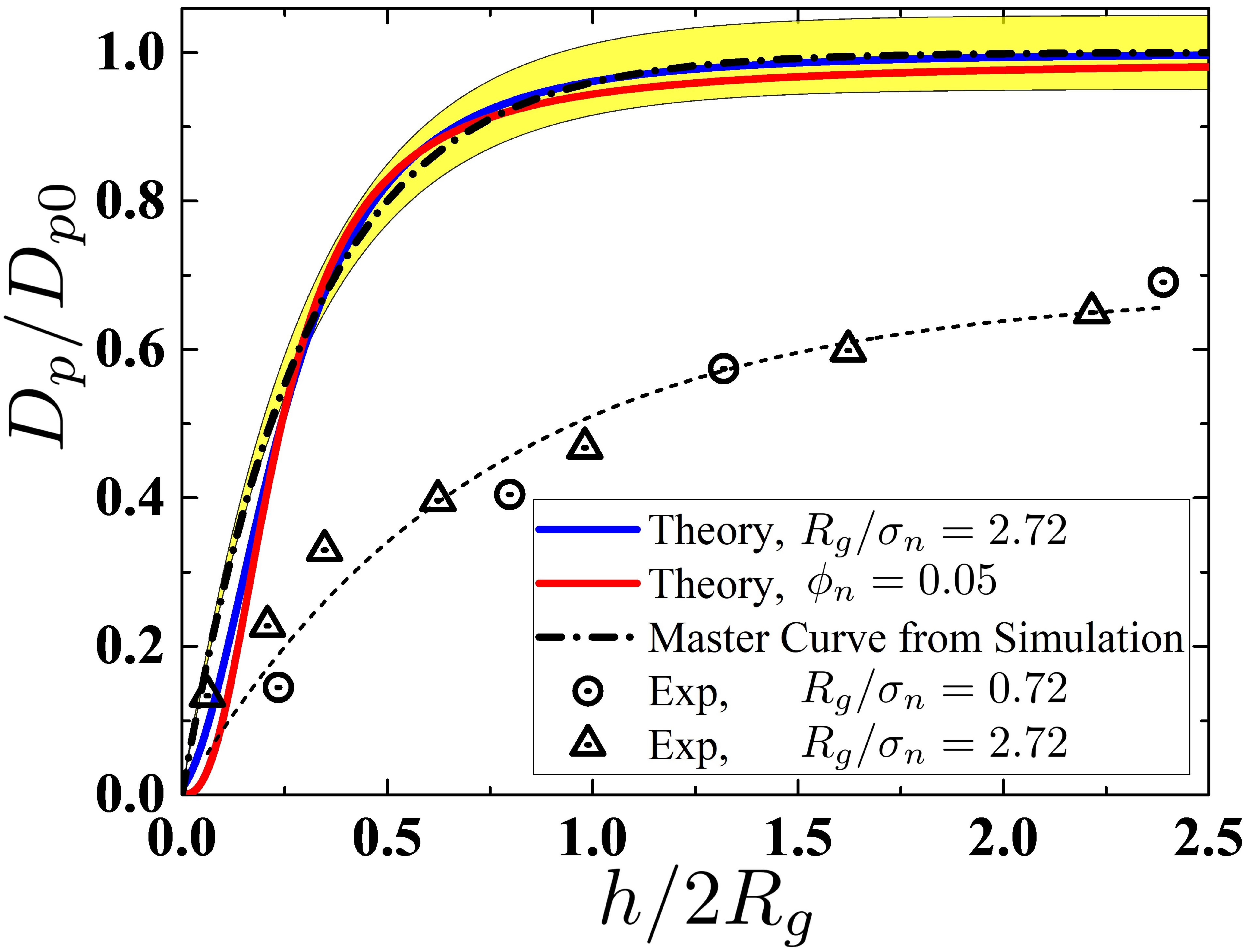}
	\caption{Normalized polymer diffusion versus ID normalized by the radius of gyration at fixed size ratio of chain length to NP diameter $R_g/\sigma_n=2.72$ (blue line) and fixed $\phi_n=0.05$ (red line), respectively.   The experimental results for two polymer-NP size ratios (symbols, ref. \citenum{Gam2011}) and a universal curve obtained from simulation (dashed-dotted line, ref. \citenum{Sorichetti2018}) are plotted.  The yellow shaded area represents $y=1-exp(-3.3x)$ with 5\% error. The dashed line is guided to the eye.}
	\label{fig:4}
\end{figure}

The NP-induced confinement effect is also characterized via the so-called average ID.  For random distribution of NPs in 3D, this quantity can be well approximated as $h(\phi_n)=\sigma_n[(\phi_{RCP}/\phi_n)^{1/3}-1]$, where $\phi_{RCP}=0.64$ is random closed packing fraction for hard spheres.  We now quantitatively calculate the dependence of polymer diffusion on the ID normalized by the twice the radius of gyration, $h/2R_g$, and compare with recent simulations and experiments data.  Fig. \ref{fig:4} shows the curves of $D_r$-$h$ obtained by changing $\phi_n$ at fixed $R_g$ (blue line) and changing $R_g$ at fixed $\phi_n$ (red line), respectively.  In general, $D_r$ drastically decreases as the $h/2R_g$ falls to zero and recovers to 1 for large normalized ID, corresponding to NP-induced slowdown of diffusion and Rouse diffusion in NP-free solution, respectively.  The dashed-dotted line in Fig. \ref{fig:4} indicates a universal curve found in the above mentioned simulation, \cite{Sorichetti2018} which shows excellent agreement with our theoretical predictions.  It can be described by an empirical expression,
\begin{equation}\label{Eq20}
	\frac{D_p}{D_{p0}}=1-exp\Big(-\frac{\alpha h}{2R_g}\Big).
\end{equation}
where $\alpha$ is fitting parameter.  The reciprocal of the dimensionless parameter, $1/\alpha$, represents a range of action of NP concentration.  The motion of polymer was remarkably inhibited when $h/2R_g<1/\alpha$.  The diffusion tends to show bulk behavior at $h/2R_g\gg 1/\alpha$.

The yellow shaded area in Fig. \ref{fig:4} shows the $D_r$-$h$ curve predicted by eqn (\ref{Eq20}) at $\alpha=3.3$ with relative error 5\%.  The empirical expression agrees well with simulation and our theoretical predictions.  Besides, we consider the presence of polymer-polymer interaction and weak NP-polymer attraction in the simulation does not greatly affect the universal curve of $D_r$-$h$.  Further, the ERD experiment shows that the data for $D_r$ as a function of $h/2R_g$ can collapse to a universal curve regardless of chain length (the symbols of Fig. \ref{fig:4}).\cite{Gam2011}  But the range of action is much greater than that predicted by theory and simulation, perhaps resulting from some kind of long-range interaction in the PNCs.  Based on current general perception, despite some universal relations are reported, \cite{Gam2011,Choi2013,Sorichetti2018} polymer dynamics depend strongly on the details in various PNCs materials, e.g., temperature and entanglement. \cite{Li2014,Composto2016}

\subsection{Polymer length effect}
\begin{figure*}[t]
	\includegraphics[width=.98\textwidth]{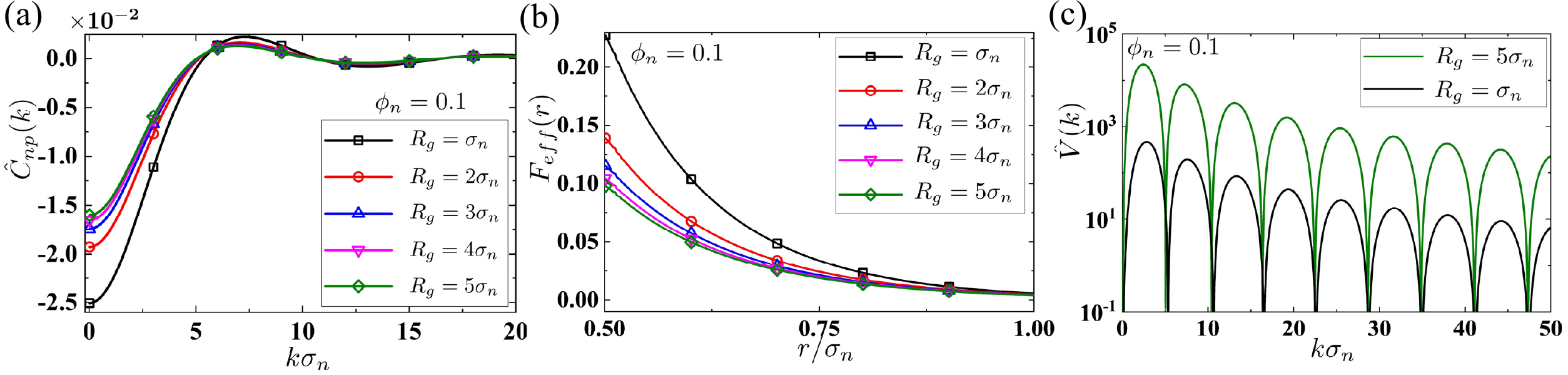}
	\caption{(a) Plots of DCF versus wavevector for several chain lengths at $\phi_n=0.1$.  (b) segment-NP distance dependence of effective force for several chain lengths. (c) Static part of force-force vertex function for two chain lengths, $R_g=5\sigma_n$ and $R_g=\sigma_n$.}
	\label{fig:5}
\end{figure*}

We now turn to focus on the polymer length effect on structure and dynamics.  Fig. \ref{fig:5}a presents the results for the DCF as a function of nondimensional wavevector $k\sigma_n$ for various polymer lengths at fixed NP volume fraction.  In contrast to the effect of NP volume fraction, the radius of gyration induces a milder decrease in the magnitude of the DCF.  The fall of the long-wavelength fluctuation $|\hat{C}(k\to 0)|$ is within 40\% when the size ratio increases from $R_g=\sigma_n$ to $5\sigma_n$.  Our calculation also reveals that chain length leads to a decrease of the effective force as shown in Fig. \ref{fig:5}b. The changing trend is consistent with that of DCF. The reduction in the DCF corresponds to a more disordered liquid structure on segment scale as chain length increases.

To further understand the role of polymer length on polymer-NP interaction and then the long-time resistance, we analyze the wavevector-dependent static vertex functions for two different size ratios, $R_g=\sigma_n$ and $R_g=5\sigma_n$.  As shown in Fig. \ref{fig:5}c, the amplitude of the vertex oscillates with $k\sigma_n$ and decays to zero for infinite $k\sigma_n$.  In contrast to the effect of NP volume fraction (Fig. \ref{fig:1}c), peak position in the vertex function keeps almost unchanged for all chain lengths studied.  It implies that chain length cannot change the average nearest-neighbor segment-NP distance.  Meanwhile, Fig. \ref{fig:5}c shows a pronounced increase in the amplitude and enveloping area of the vertex function with increasing chain length.  Hence, polymer length has a remarkable contribution on the polymer CM level through $N_p$ and $\hat{\omega}_p$ in eqn (\ref{Eq14}) into the vertex and thus into the resistance.   

\begin{figure}[!b]
	\includegraphics[width=.48\textwidth]{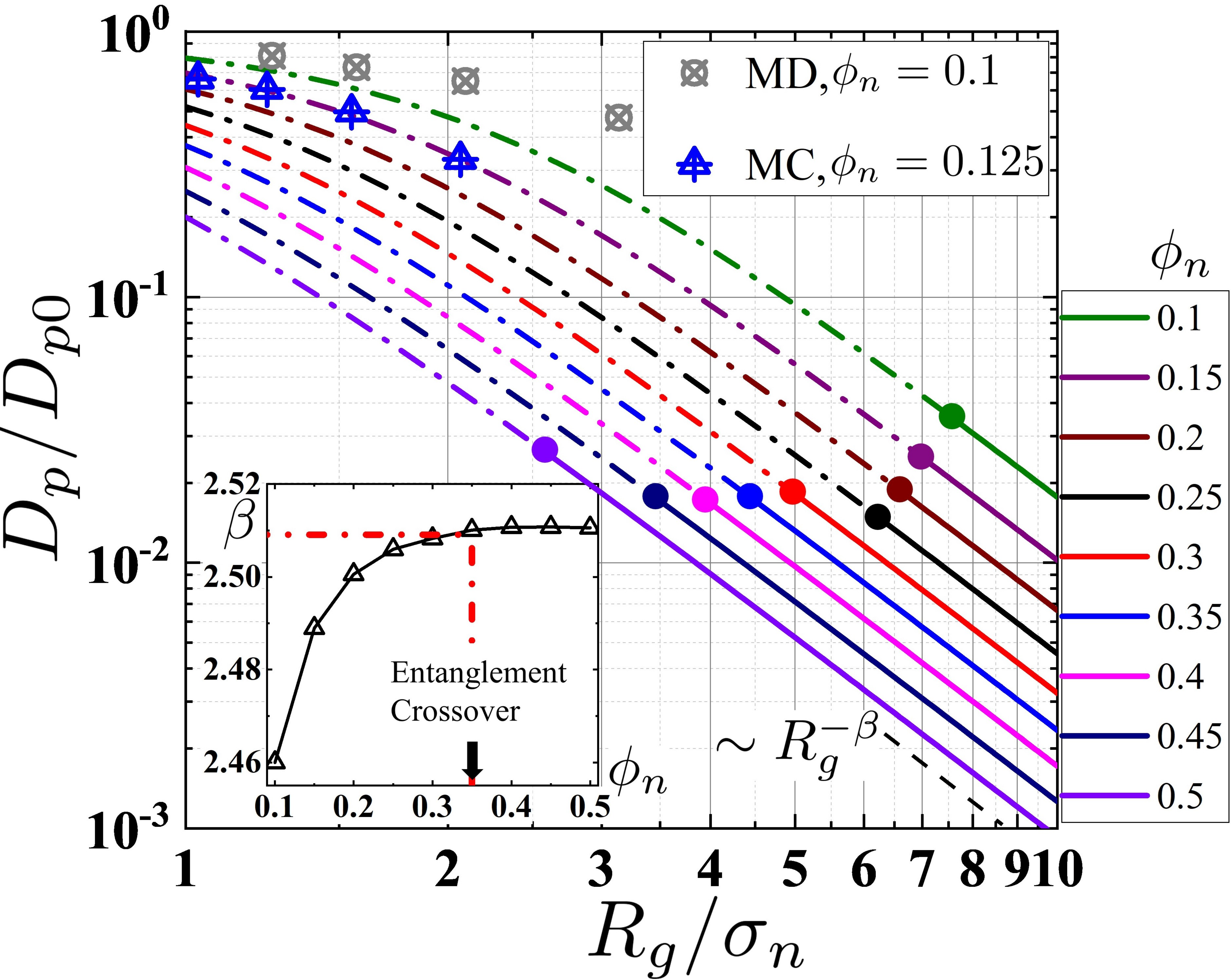}
	\caption{Normalized polymer diffusion versus the size ratio of the radius of gyration to NP diameter
		for several NP volume fractions, from $\phi_n=0.1$ (top) to $\phi_n=0.5$ (bottom). The solid lines represent the regime of power-law behavior, $D_p/D_{p0}\sim R_g^{-\beta}$.  Data from molecular dynamics simulation (ref. \citenum{Sorichetti2018}) and Monte Carlo simulation (ref. \citenum{Luo2019}) is plotted, respectively (symbols).   The solid circles mark the power-law crossover points.  Inset: The scaling exponent for several $\phi_n$.  The dashed-dotted red lines mark the $R_g$-scaling exponent at entanglement crossover NP volume fraction predicted by experiment (ref. \citenum{Richter2011}), $\beta(\phi_{nc}=0.35)\approx 2.51$.}
	\label{fig:6}
\end{figure}

To investigate the effect of polymer length on the CM dynamics, we plot the normalized polymer diffusion as a function of the polymer-NP size ratio, $R_g/\sigma_n$, for different NP volume fractions, from $\phi_n=0.1$ to $0.5$, in Fig. \ref{fig:6}.  The NP-induced reduction in the diffusion of long chain is more remarkable than that of short chain and the normalized diffusion coefficient monotonically decreases with $R_g/\sigma_n$.  More importantly, under long chain and/or high NP concentration, polymer diffusion exhibits a power-law decay, $D_r=D_p/D_{p0}\sim R_g^{-\beta}$, as illustrated in the solid lines of Fig. \ref{fig:6}.  The intersection between the dashed-pointed line and the solid line (solid circles in Fig. \ref{fig:6}), represents the crossover to power-law regime, which locates at smaller $R_g/\sigma_n$ for larger $\phi_n$.  By fitting, the $R_g$-scaling exponent is predicted as $\beta\approx 2.46$ at $\phi_n=0.1$, and tends to $\beta\approx 2.51$ at $\phi_n>0.35$ (inset of Fig. \ref{fig:6}).  

In polymer physics, the scaling relation between polymer diffusion and segment number is given by $D_p\sim N_p^{-X}$, where the $N_p$-scaling exponent $X$ depends on different surrounding environments. \cite{McLeish2002}  Considering the radius of gyration $R_g^2=N_p\sigma_p^2/6$ for Gaussian chain and diffusion constant $D_{p0} \sim N_p^{-1}$ for Rouse model, the relation of the scaling exponents between $X$ and $\beta$ is obtained as $X=\beta/2+1$.  Therefore, the prediction of our microscopic theory for the $N_p$-scaling exponent of diffusion is obtained as $X_{mix}\approx 2.230-2.225$.

In the phenomenological Doi-Edwards reptation model for entangled polymer melts, single chain is considered to be confined in an existing virtual tube formed by surrounding polymers. \cite{Doi1988}  The chain cannot move transversely across this tube due to the spatial confinement.  The diffusion and relaxation of the polymer are via reptation motion of two ends of the chain under a  fluctuating frictional force $f_{tube}$.  According to the Doi-Edwards model, the  fluctuating frictional force scales with the polymer segment number as $f_{tube}\sim {N_p}^{\gamma}$ and is proportional to the duration time moving out of the existing tube, $\tau_{tube}\sim f_{tube} L^2$, where $L$ is chain contour length and thus $L\sim N_p$.  The relation between polymer diffusion and segment number is obtained by $D_p\sim R_g^2/\tau_{tube}\sim N_p^{-1-\gamma}$.  The phenomenological model predicts the exponent $\gamma=1$ and thus the $N_p$-scaling exponent $X_{melt}=2$.  In the polymer-NP mixtures, our calculation for the scaling exponent $X_{mix}$ corresponds to $\gamma\approx 1.230-1.225$.  Essentially, the scaling exponent $\gamma$ reflects the frictional effect arising from the fluctuating force in an existing tube formed by surrounding polymers or NPs.  We believe that the exponent can characterize chain motion in various complicated environments.

In small-angle neutron scattering experiment for a silica-poly(ethylene-propylene) mixture with nonattractive polymer-NP interaction, two types of entanglement in the polymer-NP mixtures are found, chain entanglement and NP entanglement, dominating chain dynamics at low and high NP concentration, respectively. \cite{Richter2011}  When NP volume fraction increases by $\phi_n\approx 0.35$, chain dynamics exhibits a crossover from chain entanglement to NP entanglement.  In this article, our theory predicts the $N_p$-scaling exponent $X_{mix,c}=\beta/2+1\approx 2.25$ at $\phi_{nc}=0.35$ and remains almost constant at higher $\phi_n$ as shown in the inset of Fig. \ref{fig:6}.  In the coarse-grained molecular simulations, qualitative decrease of the normalized diffusion with the size ratio $R_g/\sigma_n$ is indeed observed (symbols in Fig. \ref{fig:6}). \cite{Sorichetti2018,Luo2019}  However, the exact value of $\beta$ cannot be determined due to lack of enough data in the situation of long chain or dense NPs.  A more precise measurement for the $N_p$-scaling exponent of diffusion is called for in future simulations and experiments. 
\label{sec:4}
\section{Summary and discussion}\label{sec:4}
We have constructed a first-principle theory for polymer diffusion in polymer-NP mixtures with the system-specific equilibrium structures as input.  In a minimal polymer-NP mixture model neglecting polymer-polymer interaction, an analytical expression for the CM diffusion of polymer is derived based on a combination of generalized Langevin equation, mode-coupling theory, and polymer physics.  The resistance coefficient for the CM of polymer  is expressed as an integral of time and wavevector for the length-scale-dependent vertex function, determined by equilibrium structure correlations, and dynamic density correlation propagators for tagged polymer and NPs.  The theory predicts the slowdown of nonadsorbing polymer due to fluctuating frictional force exerted by surrounding NPs.  It well captures the results from simulation and experiment studies including the dependence of polymer CM diffusion on NP volume fraction and normalized average ID.  Furthermore, numerical calculation for long chains and/or dense NPs reveals that polymer diffusion has a power-law decay as chain length, $D_p\sim N_p^{-X}$ with $X\approx 2.23$, which marks the emergence of NP-induced entanglement-like motion. 

In the theoretical framework, three major approximations are made. They could be improved by multiple avenues.  (i) Gaussian thread chain with no thickness is used to model the flexible and nonadsorbing polymer, in which nonlocal interaction between segments is neglected and the distribution of segments is statistically independent.  For more realistic polymer-NP mixtures with strong non-Gaussian behavior and complex polymer-NP interactions, a newly developed self-consistent method combining PRISM theory and Monte Carlo simulations can be applied to calculate the relevant structural properties as input. \cite{Jayaraman2018,Jayaraman2011} (ii) The resistance arising from polymer-polymer interaction is neglected in the vertex function. The explicit expression depending on static structure has been provided by the summation in eqn (\ref{Eq12}).  The term is indispensable and should be added when describing polymer diffusing in semi-dilute or dense solution, in particular in the entangled regime. (iii) The RPA and vineyard approximation adopted in the density correlation propagators correspond to the cumulant expansion up to 2nd order in displacement,\cite{Rahman1966} which is accurate to Rouse chain and Gaussian distribution.  In a self-consistent generalized Langevin equation approach, the equation for density correlation propagators can be constructed based on MCT.\cite{Yamamoto2015}  The approach is expected to provide a quantitative improvement for the approximation. 

Besides, some simulations about NP concentration in entangled polymer melts show that the NP-segment, $\sigma_n/\sigma_p$, plays important role in polymer dynamics.  For instance, Kalathi \emph{et al}. reported about 40 percent of the variation in polymer diffusion when $\sigma_n/\sigma_p=1-15$ at fixed chain length. \cite{Kalathi2014}  In this work, Gaussian thread-PRISM theory, as a Edwards-like field theory, reduces the excluded volume of segment to point-like sites ($\sigma_p\to 0$).  Thus, the minimal mixture model focuses only on the effect of $R_g/\sigma_n$ on the polymer diffusion at fixed $\sigma_n/\sigma_p$.  More realistic intramolecular structure factor with finite excluded volume of segments, such as discrete semiflexible worm-like chain, \cite{Zhang2016} should help elucidate the role of the NP-segment size ratio.  

More broadly, our tractable theoretical framework provides a foundation for various open problems about polymer dynamics in PNCs materials, such as (i) attractive polymer-NP interaction, (ii) polymer with rigidity or more complicated internal structure, such as single-chain NPs and ring polymers \cite{Moreno2020,Schweizer2020} and (iii) glassy dynamics around NP-polymer interface \cite{Napolitano2017} and activated hopping motion in dense NPs. \cite{Dell2014}  Theoretical work is ongoing in all these directions.  

\section*{Conflicts of interest}
There are no conflicts to declare

\section*{Appendix A: Explicit expressions for functions in DCF}
In the expression of the DCF (eqn (\ref{Eq9})), $\hat{Q}_{np}$, $\hat{Q}_{nn}$, $u_b$ and $v_b$ are functions of $R_g$ and $\sigma_n$.  Their derivations in detail can be found in ref. \citenum{Hansen2013,Fuchs2001}.  Here we give explicit expressions for completeness.  

$\hat{Q}_{np}$, $\hat{Q}_{nn}$ related to the real-space functions are given by three-dimensional spherically symmetric Fourier transform,
\begin{equation}
\begin{aligned}
\hat{Q}_{np}(k)& = 2\pi\int_{-1/2}^{1/2}dre^{iqr}Q_{np}(r)\\
\hat{Q}_{nn}(k)& = 2\pi\int_{0}^{1}dre^{iqr}Q_{nn}(r)
\end{aligned}
\end{equation}
where $Q_{np}(r)=0$ and $Q_{nn}(r)=0$ elsewhere.  $Q_{nn}(r)$, resulting from the correlation between hard spheres, can be found in the standard textbook, \cite{Hansen2013}
\begin{equation}
Q_{nn}(r) = \frac{A}{2}(r^2-1)+B(r-1)
\end{equation}
According to PY hard sphere solution, the coefficients $A$ and $B$ can be given by
\begin{equation}
\begin{aligned}
A=\rho_n\frac{1+2\phi_n}{(1-\phi_n)^2}, B=\rho_n\frac{-3\phi_n}{2(1-\phi_n)^2}
\end{aligned}
\end{equation}
$Q_{np}(r)$ can be obtained in ref. \citenum{Fuchs2001},
\begin{equation}
Q_{np}(r) = \frac{a}{2}(r^2-\frac{1}{4})+b(r-\frac{1}{2})\tag{A4}
\end{equation}
where the coefficients $a$ and $b$ can be written as
\begin{equation}
\begin{aligned}
a=\frac{1-\phi_n(1-6\lambda-6\xi_0)}{(1-\phi_n)^2\xi_0^2}, b=\frac{\lambda+\xi_0}{(1-\phi_n)\xi_0^2}
\end{aligned}
\end{equation}
The explicit forms of $u_b$, $v_b$ are as follows
\begin{equation}
\begin{aligned}
u_b=&-(\lambda+\xi_0) \times\frac{\xi_0-\phi_n\xi_0+\lambda[1+2\xi_0-\phi_n(1-4\xi_0)]}{(1-\phi_n)^2\xi_0^2}\\
v_b=&\frac{\lambda+\xi_0}{-\xi_0+\phi_n\xi_0}
\end{aligned}
\end{equation}
\label{A}
\section*{Appendix B: Derivation for force-force correlation function}
To further derive the force-force correlation function in eqn (\ref{Eq11}), we follow the standard steps of naive-MCT based on projected operator technique and the mode-coupling factorization approximation. \cite{Gotze2008} Here, we summarize main steps and results.  

In naive-MCT, The force-force correlation function is approximated as 
\begin{equation}
\begin{aligned}
\langle \bm{F}_{CM}(0)\cdot \bm{F}_{CM}(t)\rangle =& \langle \bm{F}_{CM}(0)e^{\Omega t} \bm{F}_{CM}(0)\rangle \\
\cong &\langle \bar{\mathcal{P}}_2 \bm{F}_{CM}(0)e^{\Omega t} \bar{\mathcal{P}}_2\bm{F}_{CM}(0)\rangle
\end{aligned}
\end{equation}
Here, the first approximation is that a operator $\bar{\mathcal{P}}_2$ projects the real fluctuating force for the center of mass of polymer onto slow modes, which are assumed to dominate the long-time dynamics.  In the mixture of chain and sphere, the slow mode is usually chosen as a bilinear product of tagged polymer density fluctuation and total collective density fluctuation, 
\begin{equation}
\hat{b}_s(\bm{k},\bm{k}')= \delta\hat{p}_{T,p}(\bm{k}) \delta\hat{c}_s(\bm{k}')
\end{equation}
with
\begin{equation}\label{EqB3}
	\begin{aligned}
		\delta\hat{p}_{T,p}(\bm{k}) = \sum_{\beta}^{N_p}e^{i\bm{k}\cdot\bm{r}_{T,p}^{\beta}}, 
 \delta\hat{c}_s(\bm{k}') = \sum_j^{n_s}\sum_{\alpha}^{N_s}e^{i\bm{k}'\cdot\bm{r}_{j,s}^{\alpha}}
	\end{aligned}
\end{equation}
where $\bm{r}_{T,p}^{\beta}$ indicates the $\beta$th monomer position of tagged polymer $p$ and $\bm{r}_{j,s}^{\alpha}$ indicates the $\alpha$th site position of $j$th molecule belonging to species type $s$.  $N_p$ is the monomer number in tagged polymer, $n_s$ is the molecule number of species type $s$ and $N_s$ is the site number in a molecule belonging to type $s$.  Hence, the projection operator is constructed as
\begin{equation}
\begin{aligned}
\bar{\mathcal{P}}_2=\frac{V^4}{(2\pi)^{12}}&\int d\bm{k}\int d\bm{k}'\int d\bm{k}''\int d\bm{k}'''\\
&\times\sum_u\sum_v \frac{\hat{b}_u(\bm{k},\bm{k}')}{\hat{B}_{uv}(\bm{k},\bm{k}',\bm{k}'',\bm{k}''')}\langle \hat{b}_v(\bm{k}'',\bm{k}''')...\rangle
\end{aligned}
\end{equation}
where the denominator $\hat{B}_{\mu\nu}$ is the normalization factor, which can be written as 
\begin{equation}
\begin{aligned}
\hat{B}_{uv}(\bm{k},\bm{k}',\bm{k}'',\bm{k}''')&\equiv \langle \hat{b}_u(-\bm{k},-\bm{k}') \hat{b}_v(\bm{k}'',\bm{k}''')\rangle\\
&=\langle \delta\hat{p}_u(-\bm{k})\delta\hat{c}_u(-\bm{k}')\delta\hat{p}_v(\bm{k}'')\delta\hat{c}_v(\bm{k}''')\rangle \\
&\cong \langle\delta\hat{p}_u(-\bm{k}) \delta\hat{p}_v(\bm{k}'')\rangle \langle\delta\hat{c}_u(-\bm{k}')\delta\hat{c}_v(\bm{k}''')\rangle
\end{aligned} 
\end{equation}
The third line in the above equation is obtained using MCT factorization method as the second approximation. \cite{Gotze2008}

Substituting the expressions of density fields given in eqn (\ref{EqB3}) into the above equation, the normalization factor is reduced as
\begin{equation}
\begin{aligned}
	\hat{B}_{uv}(\bm{k},\bm{k}',\bm{k}'',\bm{k}''')&\cong (2\pi)^6N_p\sqrt{n_uN_un_vN_v} \\
&\times\hat{S}_{uv}(\bm{k}')\hat{\omega}_p(\bm{k})V^{-2}\delta(\bm{k}-\bm{k}'')\delta(\bm{k}'-\bm{k}''')
\end{aligned} 
\end{equation}
Therefore, the projection operator is written as
\begin{equation}
\begin{aligned}
\bar{\mathcal{P}}_2=\frac{V^2}{(2\pi)^6 N_u} &\int d\bm{k} \int d\bm{k}'\sum_u\sum_v\hat{b}_u(\bm{k},\bm{k}')\\
&\times\sqrt{n_uN_un_vN_v}\hat{S}^{-1}_{uv}(\bm{k})\hat{\omega}^{-1}_u(\bm{k})\langle \hat{b}_v(\bm{k},\bm{k}')...\rangle
\end{aligned} 
\end{equation}
where the summations are over all site types in system.  Using the projection operator and following the derivation of naive-MCT in the rod-NP mixture, \cite{Jadrich2012}  the final expression in eqn (\ref{Eq12}) can be obtained. 
\label{B}

%

\section*{Acknowledgements}
This work is supported by National Natural Science Foundation of China (No.11904320, No.11847115, and No.11804085), Natural Science Foundation of Zhejiang Province (No.LQ18B040002) and Fundamental Research Funds of Zhejiang Sci-Tech University (No.18062243-Y).




\bibliography{PNCs} 
\bibliographystyle{rsc} 

\end{document}